# Relational Lattice Axioms

MARSHALL SPIGHT
Marshall.Spight@gmail.com
VADIM TROPASHKO
Vadim.Tropashko@orcl.com

___________________________________________________________________

Relational lattice is a formal mathematical model for Relational algebra. It reduces the set of six classic relational algebra operators to two: natural join and inner union. We continue to investigate Relational lattice properties with emphasis onto axiomatic definition. New results include additional axioms, equational definition for set difference (more generally anti-join), and case study demonstrating application of the relational lattice theory for query transformations.

___________________________________________________________________

## 1. INTRODUCTION

Classic Relational Algebra employs six basic operations: projection, restriction, join, union, difference, and renaming. Relational Lattice [1,2] represents them in terms of two mathematically attractive binary operations: *natural join* and *inner union*. By abuse of set notation we define these operations as follows:

- natural join

    $A(x,y) \wedge B(y,z) \overset{def}{=} \{(x,y,z) \mid (x,y) \in A \wedge (y,z) \in B\}$

- inner union

    $A(x,y) \vee B(y,z) \overset{def}{=} \{(y) \mid \exists x\, (x,y) \in A \vee \exists z\, (y,z) \in B\}$

It has been demonstrated that these operations satisfy lattice axioms, hence the notation change in favor of standard mathematical symbols for lattice meet and join. Please note that certain terminological incompatibility still remains, and we call relational operations (natural) *join* and (inner) *union* versus correspondingly lattice *meet*, and *join*. The fact that *join* term is used in both worlds to denote the exactly the opposite operation is unfortunate artifact of Relational Model legacy, which we still would like to keep.

We'll employ *Prover9* [3] as our primary theorem prover tool. Consequently, we'll use symbols ∧ (caret) for join, and ∨ (letter *v*) for union. Also, in order to avoid *Mace4* interpretation of constants, we have chosen to prefix relations 00, 01, 10 and 11 with letter R.

## 2. EXAMPLE

Relational lattices can be generated in fairly straightforward way. We start with several basic relations and generate a closure by calculating all the possible joins and unions. For example, suppose we have five relations A={(x=1)}, B={(x=1),(x=2)}, C={(y=b)}, D={(y=a),(y=b)}, and R00, where the last relation is an empty relation with empty header. Then by joining A and D we'll generate new relation {(x=1,y=a),(x=1,y=b)} which we add to the lattice. Continuing this process we'll produce the lattice on fig.1.

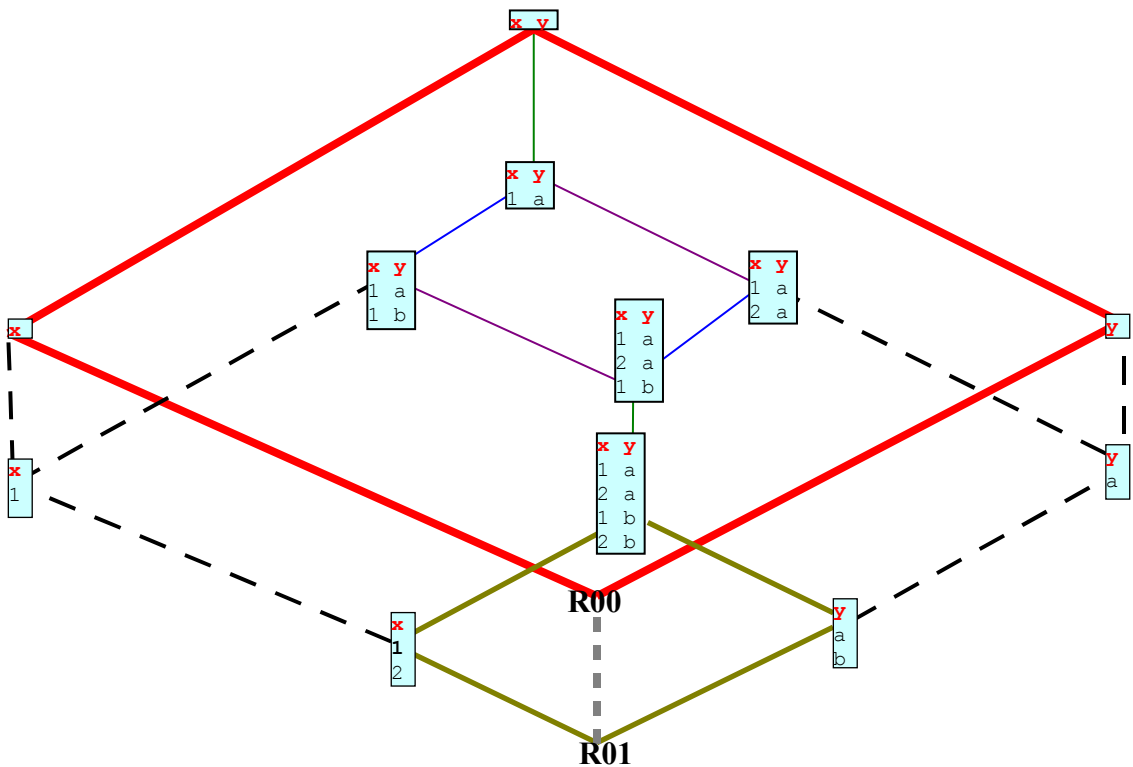

Figure 1. The lattice of relations generated by five relations A={(x=1)}, B={(x=1),(x=2)}, C={(y=a)}, D={(y=a),(y=b)}, and R00. The lattice bottom element is R01 -- the relation with empty header and nonempty content[1].

## 3. FUNDAMENTAL AXIOMS

In axiomatic description relation variables don't assumed to have any structure, other than the one specified by lattice operations. The brackets are used strictly for algebraic purpose of grouping operations. We would never be forced to specify columns explicitly (like in the introduction section in case of the relations A(x,y) and B(y,z) ); consequently, comma symbol would never make an appearance. Therefore, the case for variable is no longer important, and we can switch to lower case letters x, y, z, etc for relation variables.

---

[1]Relations R00 and R01 are known in database literature under colorful names of *Dee* and *Dum*

We partition Relational axioms into several groups. In this section we list standard lattice axioms (SLA)

    x ^ y = y ^ x.[2]

    (x ^ y) ^ z = x ^ (y ^ z).

    x ^ (x v y) = x.

    x v y = y v x.

    (x v y) v z = x v (y v z).

    x v (x ^ y) = x.

together with *Fundamental Decomposition Identity* (FDA)

    x = (x ^ R00) v (x ^ R11).

where R11 is the universal relation. In the lattice example from the previous section (Fig.1) for all practical purposes R11 can be identified with the relation {(x=1,y=a),(x=1,y=b),(x=2,y=a),(x=2,y=b)}.

Informally, FDA asserts that any relation is a union of relation's content and header. The dual of FDA

    ~~x = (x v R00) ^ (x v R11).~~

is invalid axiom. In the lattice example from the previous section let

    x $\stackrel{\text{def}}{=}$ {(x=1,y=a),(x=1,y=b),(x=2,y=a)}

then x v R00 evaluates to R01, and x v R11 to R11. Consequently, R01 ^ R11 = R11 ≠ x.

If both sides of the dual of FDA are joined with R00, then we obtain a weaker axiom which (for the lack of better name) we call FDA$^{-1}$

    **R00 ^ (x v R11) = x ^ R00.**

Unlike previously considered dual of FDA, FDA$^{-1}$ is a legitimate axiom.

It is easy to see that FDA is independent from SLA. Choose R00 = R11 to be an arbitrary element in a lattice. Then, x ^ R00 = x ^ R11 therefore, (x ^ R00) v (x ^ R11) evaluates to x ^ R00 which would not reduce to x if R00 ≰ x. To witness that FDA$^{-1}$ is independent from the SLA+FDA take the standard 2-element boolean algebra {1,0}, 0 ^ 1 = 1, 0 v 1 = 0, and assume R00 = R11 = 0. Then, the right side of the FDA$^{-1}$ reduces to 0 by absorption. Choose x = 1 on the right side.

Speaking universal algebra language, so far we have introduced two binary operations – join and union, and two constants -- R00 and R11 (which can be viewed as 0-ary operations). What about two more constants, which were introduced in [2]: R10 (lattice top) and R01 (lattice bottom)? They were defined to satisfy the equations

---

[2] Dots are essential part of Prover9 syntax

$$x \wedge R01 = x.$$
$$x \vee R10 = x.$$

Let's substitute $x = R01$ into FDA. It immediately follows that
$$R00 \vee R11 = R01.$$
Likewise, one can prove
$$R00 \wedge R11 = R10.$$
These two theorems exhaust all the interesting propositions about R10 and R01. For that reason we have chosen not to mention R10 and R01 (and their defining axioms) at all.

$FDA^{-1}$ has two more cousins which can be derived from $SLA+FDA+FDA^{-1}$:
$$R11 \vee (x \wedge R00) = x \vee R11.$$
$$R00 \vee (x \wedge R11) = x \vee R00.$$
In appendix A we list assertions and goal, so that user could be able to reproduce Prover9 generated proof.

We complete the section with one more theorem
$$R00 \wedge x = R00 \wedge y \iff R11 \vee x = R11 \vee y.$$
see appendix B. Informally, given a condition onto relation headers, it can be rewritten in terms of domains and vice versa. This theorem is widely leveraged in the next section.

4. DISTRIBUTIVITY AXIOMS

It is easy to spot the "pentagon" $N_5$ sublattices on Fig.1. Therefore, Relational lattice in general is not distributive. $SLA+FDA+FDA^{-1}$ axiom system, however, possesses some rudimentary distributivity properties such as
$$(R00 \wedge x) \vee (R00 \wedge y) = R00 \wedge (x \vee y).$$
-- see appendix C. By associativity and absorption we also have
$$(R00 \wedge x) \wedge (R00 \wedge y) = R00 \wedge (x \wedge y).$$
These assertions establish that the mapping $x \to R00 \wedge x$ is a *lattice homomorphism* (LH^). The dual mapping $x \to R00 \vee x$ is *not* a homomorphism.

Likewise, the mapping $x \to R11 \vee x$ is a homomorphism (LHv). The dual mapping $x \to R11 \wedge x$ is not a homomorphism, either.

Relational lattice honors conditional distributivity laws that are stronger than aforementioned LH^ and LHv theorems. In previous work ([2]) we discovered two distributivity criteria and proved them within set theoretic framework. The first criteria for distributivity of join over union translates into the following implication

$$(R00 \wedge (x \vee y) = R00 \wedge (x \vee z)) \rightarrow (x \wedge (y \vee z) = (x \wedge y) \vee (x \wedge z)).$$

This axiom, which we'll refer to as SDC, is independent from SLA+FDA+FDA$^{-1}$; please refer to appendix D for counterexample. The dual criteria for distributivity of union over join

$$(R00 \wedge (x \vee y) = R00 \wedge (x \vee z)) \& (R00 \wedge (x \vee z) = R00 \wedge (y \vee z)) \rightarrow$$
$$(x \vee (y \wedge z) = (x \vee y) \wedge (x \vee z)).$$

is a theorem in the SLA+FDA+FDA$^{-1}$+SDC system; please refer to appendix E.

These (and the assertions that follows) can be rewritten in terms of condition over domains, rather than relation headers based upon the theorem that we derived at the end of section 3.

Unfortunately, SLA+FDA+FDA$^{-1}$+SDC system is still too weak. Distributivity Constraint on relation Headers (DCH)

$$R00 \wedge (x \wedge (y \vee z)) = R00 \wedge ((x \wedge y) \vee (x \wedge z)).$$

is one more axiom (Appendix F). The dual assertion

$$R00 \wedge (x \vee (y \wedge z)) = R00 \wedge ((x \vee y) \wedge (x \vee z)).$$

is a theorem (Appendix G).

Closely related to DCH is the law of Distributivity of Empty Relations

$$(x \wedge R00) \wedge ((y \wedge R00) \vee (z \wedge R00)) = ((x \wedge R00) \wedge (y \wedge R00)) \vee ((x \wedge R00) \wedge (z \wedge R00)).$$
$$(x \wedge R00) \vee ((y \wedge R00) \wedge (z \wedge R00)) = ((x \wedge R00) \vee (y \wedge R00)) \wedge ((x \wedge R00) \vee (z \wedge R00)).$$

see appendix H.

## 5. DISTRIBUTIVITY IN DATE&DARWEN ALGEBRA

Date & Darwen introduced the OR operation [4] as follows

$$A(x,y) \text{ OR } B(y,z) \stackrel{def}{=} \{(x,y,z) \mid (x,y) \in A \wedge z \in Z\} \cup \{(x,y,z) \mid x \in X \wedge (y,z) \in B\}$$

It is represented in relational lattice terms as

$$x + y \stackrel{def}{=} (x \wedge (y \vee R11)) \vee (y \wedge (x \vee R11)).$$

where we found it convenient to switch the notation to the plus symbol in order to continue leveraging Prover9 system. Associativity of the OR operation and distributivity of join over the OR

$$x + (y + z) = (x + y) + z.$$
$$x \wedge (y + z) = (x \wedge y) + (x \wedge z).$$

are theorems in the SLA+FDA+FDA$^{-1}$+SDC+DCH system (Appendixes I and J). Unfortunately, we were unable to find the proof, nor counterexample for distributivity of the OR operation over join

$$x + (y \wedge z) = (x + y) \wedge (x + z).$$

## 6. ANTIJOIN

Among six basic Relational algebra operations only five are directly representable via join and union. Set difference is the exception. Yet, set difference is a special case of anti-join, which we can define equationally in relational lattice theory. Let's lower abstraction level a little, and introduce relational variables E and D which we informally associate with the familiar tables Emp(ename,deptno) and Dept(deptno) (Fig.2).

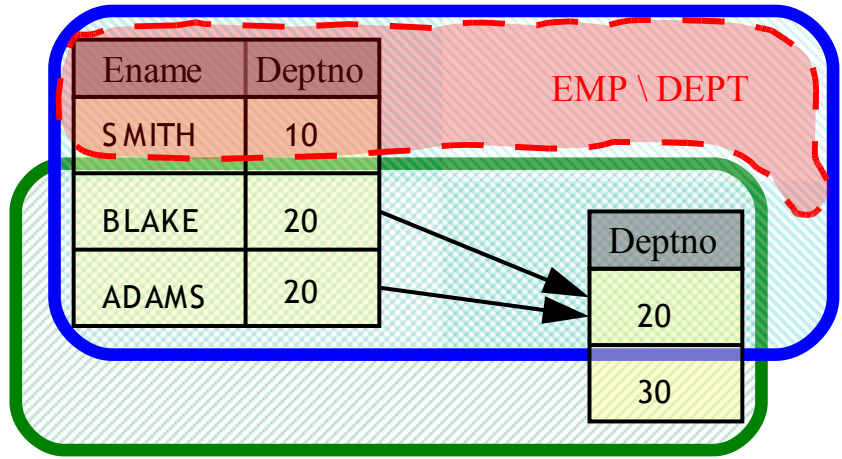

Figure 2. The record (Ename=SMITH, Deptno=10) doesn't match any record in the department table: it belongs to the set E \ D (where we extended set difference notation for anti-join).

Once again, we are trying to define anti-join/set difference equationally, following pretty much the same idea how arithmetic difference is defined (that is as a solution of the equation x + a = b in classic algebra). Therefore, let's introduce variable EmD[3] for antijoin E \ D. It satisfies the following system of equations:

$$(E \wedge D) \vee EmD = E.$$
$$(E \wedge D) \wedge EmD = (E \wedge D) \wedge R00.$$

These equations informally assert that the relation E is a disjoint union of EmD and join of E and D.

Anti-join is well defined thanks to the following *uniqueness* theorem. Given "alternative" anti-join variable EmD1

$$(E \wedge D) \vee EmD1 = E.$$
$$(E \wedge D) \wedge EmD1 = (E \wedge D) \wedge R00.$$

one can formally prove the equality

$$EmD1 = EmD.$$

---

[3]We apologize to mathematically inclined readers for introducing programming flavored variable names

(Appendix K)

## 7. CASE STUDY: REDUNDANT JOIN ELIMINATION

Eliminating redundant joins [5] is one of the "obvious" query transformations which we are going to test our method against. Assume foreign key constraint between the E and D tables from the previous section, that is the empty EmD relation. Formally,

$$\text{EmD} \lor \text{R00} = \text{R00}.$$

Consider a query where we join E and D first, and then project the result to a subset of columns of E. We define the "subset of columns of E" as a formal relation E0 that satisfies[4]

$$(E \land E0) \land R00 = E \land R00.$$

Now we can formally express our query as

$$E0 \lor (E \land D).$$

Our goal is to prove

$$E0 \lor (E \land D) = E0 \lor E.$$

Once again, we plug all the above equations into Prover9 theorem prover, and effortlessly derive the desired result (Appendix L).

## 8. ACKNOWLEDGMENTS

Jan Hidders spotted an error in the distributivity of union over join criteria, and is credited with the idea of splitting the + - associativity theorem in half.

## REFERENCES

1. TROPASHKO, V. 2005. *Relational Algebra as non-distributive lattice.* http://arxiv.org/pdf/cs.DB/0501053
2. SPIGHT, M., TROPASHKO, V. 2006. *First Steps in Relational Lattice.* http://arxiv.org/pdf/cs.DB/0603044
3. MCCUNE, W. *Prover9 and Mace4.* http://www.cs.unm.edu/~mccune/mace4/
4. DATE, C.J., DARWEN, H. 2000. *Foundation for Future Database Systems: The Third Manifesto.* Addison-Wesley.
5. WAINGOLD/LEE, A. 2008. *Why are some of the tables in my query missing from the plan?* http://optimizermagic.blogspot.com/2008/06/why-are-some-of-tables-in-my-query.html

---

[4] Note that usually we associate relation headers with empty relations, so that we may want to add one more condition

$$E0 \land R00 = E0.$$

This constraint, however, is non essential in our case study, so we'll drop it from our consideration.

## APPENDIX A. FDA$^{-1}$ COUSIN IDENTITIES

```
R11 v (x ^ R00) = x v R11.  [goal].
x ^ y = y ^ x.  [assumption].
x ^ (x v y) = x.  [assumption].
x v y = y v x.  [assumption].
x = (x ^ R00) v (x ^ R11).  [assumption].
R00 ^ (x v R11) = x ^ R00.  [assumption].

R00 v (x ^ R11) = x v R00.  [goal].
x ^ y = y ^ x.  [assumption].
x v y = y v x.  [assumption].
(x v y) v z = x v (y v z).  [assumption].
x v (x ^ y) = x.  [assumption].
x = (x ^ R00) v (x ^ R11).  [assumption].
```

## APPENDIX B. EQUIVALENCE OF CONDITIONS OVER HEADERS AND DOMAINS

```
R11 v x = R11 v y <-> R00 ^ x = R00 ^ y.  [goal].
x ^ y = y ^ x.  [assumption].
x ^ (x v y) = x.  [assumption].
x v y = y v x.  [assumption].
x = (x ^ R00) v (x ^ R11).  [assumption].
R00 ^ (x v R11) = x ^ R00.  [assumption].
```

## APPENDIX C. DISTRIBUTIVITY FOR R00 ELEMENT

```
(R00 ^ x) v (R00 ^ y) = R00 ^ (x v y).  [goal].
x ^ y = y ^ x.  [assumption].
(x ^ y) ^ z = x ^ (y ^ z).  [assumption].
x ^ (x v y) = x.  [assumption].
x v y = y v x.  [assumption].
(x v y) v z = x v (y v z).  [assumption].
x v (x ^ y) = x.  [assumption].
x = (x ^ R00) v (x ^ R11).  [assumption].
R00 ^ (x v R11) = x ^ R00.  [assumption].
```

## APPENDIX D. MODEL FOR SDC

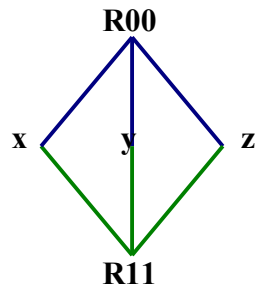

Figure 3. 5-element "diamond" lattice M3 as a model satisfying SLA+FDA+FDA$^{-1}$ but invalidating SDC axiom.

## APPENDIX E. PROOF FOR DISTRIBUTIVITY OF UNION OVER JOIN

```
R00 ^ (x v y) = R00 ^ (x v z) -> x ^ (y v z) = (x ^ y) v (x ^ z).  [assumption].
R00 ^ (x v y) = R00 ^ (x v z) & R00 ^ (x v z) = R00 ^ (y v z) -> x v (y ^ z) = (x v y) ^
(x v z).  [goal].
x ^ y = y ^ x.  [assumption].
(x ^ y) ^ z = x ^ (y ^ z).  [assumption].
x ^ (x v y) = x.  [assumption].
x v y = y v x.  [assumption].
```

```
(x v y) v z = x v (y v z).   [assumption].
x v (x ^ y) = x.   [assumption].
x = (x ^ R00) v (x ^ R11).   [assumption].
R00 ^ (x v R11) = x ^ R00.   [assumption].
```

## APPENDIX F. MODEL FOR DISTRIBUTIVITY CONSTRAINT ON RELATION HEADERS

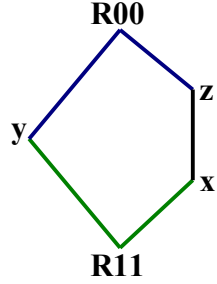

Figure 4. Familiar 5-element "pentagon" $N_5$ as a model satisfying SLA+FDA+FDA$^{-1}$+SDC but invalidating DCH axiom.

## APPENDIX G. DUAL OF DISTRIBUTIVITY CONSTRAINT ON RELATION HEADERS

```
R00 ^ (x v y) = R00 ^ (x v z) -> x ^ (y v z) = (x ^ y) v (x ^ z).   [assumption].
R00 ^ (x v (y ^ z)) = R00 ^ ((x v y) ^ (x v z)).   [goal].
x ^ y = y ^ x.   [assumption].
(x ^ y) ^ z = x ^ (y ^ z).   [assumption].
x ^ (x v y) = x.   [assumption].
x v y = y v x.   [assumption].
(x v y) v z = x v (y v z).   [assumption].
x v (x ^ y) = x.   [assumption].
R00 ^ (x ^ (y v z)) = R00 ^ ((x ^ y) v (x ^ z)).   [assumption].
```

## APPENDIX H. DISTRIBUTIVITY OF UNION OVER JOIN FOR EMPTY RELATIONS

```
R00 ^ (x v y) = R00 ^ (x v z) -> x ^ (y v z) = (x ^ y) v (x ^ z).   [assumption].
(x ^ R00) ^ ((y ^ R00) v (z ^ R00)) = ((x ^ R00) ^ (y ^ R00)) v ((x ^ R00) ^ (z ^ R00)).
[goal].
x ^ y = y ^ x.   [assumption].
(x ^ y) ^ z = x ^ (y ^ z).   [assumption].
x ^ (x v y) = x.   [assumption].
R00 ^ (x ^ (y v z)) = R00 ^ ((x ^ y) v (x ^ z)).   [assumption].
```

## APPENDIX I. ASSOCIATIVITY OF D&D *'OR'* OPERATOR

This is one of the proofs where Prover9 is pushed to its limits. We splitted the problem in half and proved

(x + y) + z = ((x ^ ((y ^ z) v R11)) v (y ^ ((x ^ z) v R11))) v (z ^ ((x ^ y) v R11)).

We also represented SDC and DCH in the domain form

(R11 v (x v y) = R11 v (x v z))  -> (x ^ (y v z) = (x ^ y) v (x ^ z)).

R11 v (x ^ (y v z)) = R11 v ((x ^ y) v (x ^ z)).

```
R11 v (x v y) = R11 v (x v z) -> x ^ (y v z) = (x ^ y) v (x ^ z).
[assumption].
(x + y) + z = ((x ^ ((y ^ z) v R11)) v (y ^ ((x ^ z) v R11))) v (z ^ ((x ^ y) v R11)).
[goal].
x ^ y = y ^ x.   [assumption].
```

```
(x ^ y) ^ z = x ^ (y ^ z).  [assumption].
x ^ (x v y) = x.  [assumption].
x v y = y v x.  [assumption].
(x v y) v z = x v (y v z).  [assumption].
x v (x ^ y) = x.  [assumption].
x = (x ^ R00) v (x ^ R11).  [assumption].
R00 ^ (x v R11) = x ^ R00.  [assumption].
R11 v (x ^ (y v z)) = R11 v ((x ^ y) v (x ^ z)).  [assumption].
x + y = (x ^ (y v R11)) v (y ^ (x v R11)).  [assumption].
```

## APPENDIX J. DISTRIBUTIVITY OF D&D *'OR'* OPERATOR

```
R11 v (x v y) = R11 v (x v z) -> x ^ (y v z) = (x ^ y) v (x ^ z).  [assumption].
x ^ (y + z) = (x ^ y) + (x ^ z).  [goal].
x ^ y = y ^ x.  [assumption].
(x ^ y) ^ z = x ^ (y ^ z).  [assumption].
x ^ (x v y) = x.  [assumption].
x v y = y v x.  [assumption].
(x v y) v z = x v (y v z).  [assumption].
x v (x ^ y) = x.  [assumption].
R11 v (x ^ (y v z)) = R11 v ((x ^ y) v (x ^ z)).  [assumption].
(x v R11) ^ (y v R11) = (x ^ y) v R11.  [assumption].
x + y = (x ^ (y v R11)) v (y ^ (x v R11)).  [assumption].
```

## APPENDIX K. ANTI-JOIN UNIQUENESS

```
R00 ^ (x v y) = R00 ^ (x v z) -> x ^ (y v z) = (x ^ y) v (x ^ z).  [assumption].
EmD1 = EmD.  [goal].
x ^ y = y ^ x.  [assumption].
(x ^ y) ^ z = x ^ (y ^ z).  [assumption].
x ^ (x v y) = x.  [assumption].
x v y = y v x.  [assumption].
(x v y) v z = x v (y v z).  [assumption].
x v (x ^ y) = x.  [assumption].
x = (x ^ R00) v (x ^ R11).  [assumption].
R00 ^ (x v R11) = x ^ R00.  [assumption].
R00 ^ (x ^ (y v z)) = R00 ^ ((x ^ y) v (x ^ z)).  [assumption].
R00 ^ (x v (y ^ z)) = R00 ^ ((x v y) ^ (x v z)).  [assumption].
(E ^ D) v EmD = E.  [assumption].
(E ^ D) ^ EmD = (E ^ D) ^ R00.  [assumption].
(E ^ D) v EmD1 = E.  [assumption].
(E ^ D) ^ EmD1 = (E ^ D) ^ R00.  [assumption].
```

## APPENDIX L. REDUNDANT JOIN ELIMINATION

```
R00 ^ (x v y) = R00 ^ (x v z) -> x ^ (y v z) = (x ^ y) v (x ^ z).  [assumption].
E0 v (E ^ D) = E0 v E.  [goal].
x ^ y = y ^ x.  [assumption].
(x ^ y) ^ z = x ^ (y ^ z).  [assumption].
x ^ (x v y) = x.  [assumption].
x v y = y v x.  [assumption].
(x v y) v z = x v (y v z).  [assumption].
x v (x ^ y) = x.  [assumption].
(E ^ D) v EmD = E.  [assumption].
(E ^ D) ^ EmD = (E ^ D) ^ R00.  [assumption].
(E ^ E0) ^ R00 = E ^ R00.  [assumption].
EmD v R00 = R00.  [assumption].
```